\documentclass[prc,twocolumn,superscriptaddress,nofootinbib]{revtex4-1}
\usepackage{newtxtext}
\usepackage[varvw,bigdelims]{newtxmath}
\usepackage[colorlinks=true,allcolors=blue]{hyperref}
\usepackage{graphicx}
\usepackage{bm}
\usepackage{amsmath}
\usepackage{cancel}
\usepackage{ulem}
\usepackage{physics}
\usepackage{xcolor}
\usepackage{orcidlink}

\newcommand{\GXNU}{\affiliation{Department of Physics, Guangxi Normal University, Guilin 541004, China}}

\newcommand{\GXZD}{\affiliation{Guangxi Key Laboratory of Nuclear Physics and Technology, Guangxi Normal University, Guilin 541004, China}}

\newcommand{\KobeU}{\affiliation{Graduate School of Maritime Sciences, Kobe University, Kobe 658-0022, Japan}}

\newcommand{\IFIC}{\affiliation{Departamento de F\'{\i}sica Te\'orica and IFIC, Centro Mixto Universidad de
Valencia-CSIC Institutos de Investigaci\'on de Paterna, Apartado 22085, 46071 Valencia, Spain}}

\begin{document}
\title{ \boldmath The role of the $f_0(1710)$ and $a_0(1710)$ resonances in the $D^0 \to \rho^0 \phi$, $\omega \phi$ decays}

\author{Natsumi Ikeno}
\email{ikeno@maritime.kobe-u.ac.jp}
\GXNU
\KobeU

\author{Wen-Hao Jia}%
\GXNU
 
\author{Wei-Hong Liang}%
\email{liangwh@gxnu.edu.cn}
\GXNU
\GXZD

\author{Eulogio Oset}
\email{oset@ific.uv.es}
\GXNU
\IFIC


\date{\today}

\begin{abstract}

We study the $D^0 \to \rho^0 \phi$, $\omega \phi$ decays which proceed in a direct mode via internal emission with equal rates. Yet, the experimental branching ratio for the $\rho^0 \phi$ mode is twice as big as that for the $\omega \phi$ mode. We find a natural explanation based on the extra indirect mechanism where $K^{*+} K^{*-}$ is produced via external emission and that channel undergoes final state interaction with other vector--vector  channels to lead to the $\rho^0 \phi$, $\omega \phi$ final states, with transition amplitudes dominated by the $a_0(1710)$ resonance, recently discovered, and $f_0(1710)$ respectively. The large coupling of the $a_0(1710)$ to the $\rho^0 \phi$ channel is mostly responsible for this large ratio of the production rates. 
\end{abstract}


\maketitle

\section{Introduction}

\begin{figure}[!b]
\centering
\includegraphics[width=0.43\textwidth]{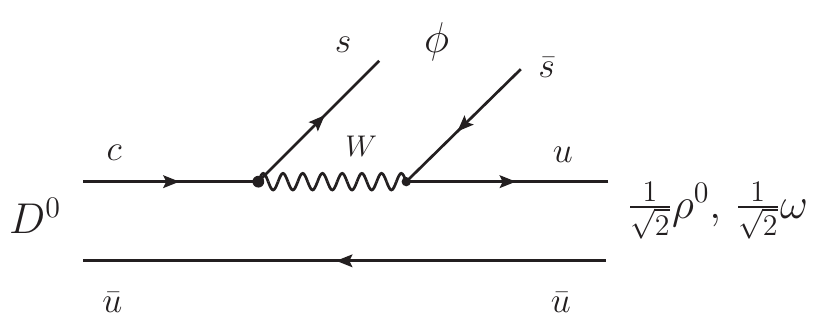}
\caption{Diagram for $D^0 \to \phi \rho^0$, $\phi \omega$ at the quark level, involving internal emission. The $W \bar{s} u$ vertex is Cabibbo suppressed.}
\label{fig:1}
\end{figure}

The branching fractions of $D^0 \to \rho^0 \phi$, $\omega \phi$ are given in the PDG~\cite{ParticleDataGroup:2024cfk} as
\begin{align}
& {\rm Br}(D^0 \to \omega \phi)  = (6.5 \pm 1) \times 10^{-4},   \label{eq:Br_omega}\\
& {\rm Br}(D^0 \to \rho^0 \phi ;\ \phi \to K^+ K^-)  = (6.9 \pm 0.6) \times  10^{-4} , \label{eq:Br_rho1}
\end{align}
which, considering that the branching ratio of the $\phi \to K^+ K^-$is $(49.1 \pm 0.5)\%$~\cite{ParticleDataGroup:2024cfk}, implies,
dividing Eq.~\eqref{eq:Br_rho1} by the $\phi \to K^+ K^-$ branching ratio and summing relative errors in quadrature,
\begin{equation}
{\rm Br}(D^0 \to \rho^0 \phi)  = (14.1 \pm 1.2) \times 10^{-4} . \label{eq:Br_rho}
\end{equation}
Note that $S$-wave dominance is assumed in both Eqs.~\eqref{eq:Br_omega} and \eqref{eq:Br_rho1} which is corroborated in the PDG for the case of Eq.~\eqref{eq:Br_rho1}.
The $D^0 \to \rho^0 \phi$ rate is about twice as big as that of $D^0 \to \omega \phi$.
 This result is surprising since the process, which is single Cabibbo suppressed, and requires internal emission~\cite{Chau:1982da} since one is producing neutral particles, provides the same decay amplitude in both cases and the phase space for decay is also about the same. 
 This can be easily seen in the topological diagram of Fig.~\ref{fig:1} at the quark level.

Since $\phi \equiv s \bar s$ and
\begin{equation}
\rho^0 = \frac{1}{\sqrt{2}} ( u \bar u - d \bar d ), ~~~~~~
\omega = \frac{1}{\sqrt{2}} ( u \bar u + d \bar d ),
\end{equation}
the overlap of $\rho^0$, $\omega$ with the $u \bar u$ component gives $\frac{1}{\sqrt{2}}$ in both cases. 
The phase space, the $\rho^0$ or $\omega$ momentum in the $D^0 \to \phi \rho^0$, $\phi \omega$ decay, only gives a $7\%$ increase in the $D^0 \to \phi \rho^0$ decay versus $D^0 \to \phi \omega$, because of the smaller $\rho$ mass, clearly far away from the factor $2.17$ ratio of the $\rho^0 \phi$, $\omega \phi$ rates of the experiment, Eqs.~\eqref{eq:Br_omega}-\eqref{eq:Br_rho}.

The rescue could come from producing other states in a primary step and letting them interact through final state interaction to lead to $ \rho^0 \phi$, $\omega \phi$. 
This mechanism could benefit from having this primary process occurring through external emission, which is color favored by a factor $N_c \equiv 3$. 
The final state interaction requires to apply dynamics of strong interaction, and indeed the weak decays are an excellent source of information on strong interaction. 
An idea of the relevance and interest in this subject can be seen in the following list, not complete, of works done along this line~\cite{Oset:2016lyh,Wang:2024gsh,Ding:2024lqk,Liu:2023jwo,Lyu:2023ppb,Luo:2022iqd,Zhu:2022guw,Wang:2022nac,Ahmed:2021oft,Wang:2021ews,Wang:2021naf,Ling:2021qzl,Wei:2021usz,Ikeno:2021kzf,Liu:2020ajv,Feng:2020jvp,Ikeno:2024fjr,Toledo:2020zxj,Duan:2020vye,Wang:2020pem,Zhang:2020rqr,Molina:2019udw,Wang:2019mph,Rui:2018mxc,Dai:2018hqb,Miyahara:2018lud,Pavao:2018wdf,Dai:2018tgo,Liang:2017ijf,Sakai:2017iqs,Xie:2017xwx,Albaladejo:2016mad,Liang:2016ydj,Xie:2016evi,Wang:2016wpc,Dias:2016gou,Chen:2015sxa,El-Bennich:2006rcn,Furman:2005xp,Dedonder:2014xpa,Furman:2011zp,Sakai:2020psu,Niecknig:2017ylb,Niecknig:2015ija,Daub:2015xja,Liang:2014tia,Xie:2014tma,Roca:2015tea,Liang:2015qva,Wang:2015pcn,Miyahara:2016yyh,Pavao:2017cpt,Xie:2018gbi,Dias:2021upl,Roca:2020lyi,Magalhaes:2011sh,Magalhaes:2015fva,Boito:2009qd,Diakonou:1989sf,Jia:2024pyb,Niecknig:2018oqn,Xie:2018rqv,Liu:2019ymi,Rui:2019yxx,Sakai:2020fjh,Ahmed:2020qkv,He:2021exv,Peng:2024ive,Liang:2023ekj,Wang:2023aza,Abreu:2023hts,He:2023plt,Zhang:2022xpf,Wang:2022xqc,Song:2022kac,Wang:2021kka}.
Detailed and pedagogical information can be seen in the review paper~\cite{Oset:2016lyh} and the PhD Thesis of Ref.~\cite{Niecknig:2018oqn}.

Since one has vector--vector ($VV$) in the final state, we search for other $VV$ states which are produced by external emission. 
Then we shall rely on the study of the $VV$ interaction in the local hidden gauge, following the work of Refs.~\cite{Molina:2008jw,Geng:2008gx,Du:2018gyn}. 
The mechanism for this reaction is discussed in the next section.

\section{Role of $VV$ interaction in the $D^0 \to \rho^0 \phi$, $\omega \phi$ decays}
\begin{figure}[!htb]
\centering
\includegraphics[width=0.43\textwidth]{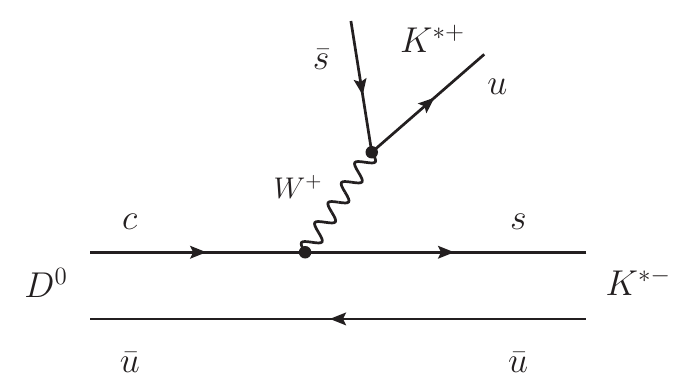}
\caption{ Mechanism of $K^{*+} K^{*-}$ production through external emission.}
\label{fig:2}
\end{figure}

We now look at external emission at the quark level and we can see that again, with a Cabibbo suppressed vertex $W^+ \bar s u$, we can produce $K^{*+} K^{*-}$ in a first step, see Fig.~\ref{fig:2}.

The $K^{*+} K^{*-}$ is not the desired  final state, but we can implement the final state interaction of $K^{*+} K^{*-} \to \rho^0 \phi$, $\omega \phi$ to produce $\rho^0 \phi$ and $\omega \phi$ at the end.
The idea of using this mechanism was already introduced in Ref. \cite{Cao:2023csx}. 
However a non-perturbative approach to this problem was used, which does not generate the $f_0(1710)$ and $a_0(1710)$ resonances. 
The study of vector-vector interaction was conducted in Ref.~\cite{Geng:2008gx} using the local hidden gauge approach and coupled channels, and it was found that the scattering matrices developed poles for the resonance $f_0(1710)$ in isospin $I=0$, and another resonance with $I=1$ with a mass around 1777~MeV. 
At the time of the prediction of this $a_0(1777)$ resonance, there was no experimental information on that state. 
The situation changed recently when three different collaborations, Babar~\cite{BaBar:2021fkz}, BESIII~\cite{BESIII:2021anf,BESIII:2022npc}, and LHCb~\cite{LHCb:2023evz} found evidence for the existence of this state. 
There is still some discrepancy about the mass of the state, but all these works find evidence for the existence of the state. 
In the PDG this resonance is cataloged as the $a_0(1710)$. In Table~\ref{tab:1} we give the information about the mass and width from these experiments.

\begin{table}[!tb]
\centering
\caption{Masses and widths for the $a_0(1710)$.}
\begin{tabular}{c| c c }
\hline 
   &~~ Mass [MeV] ~~ & ~~ $\Gamma$ [MeV]~~   \\
\hline
PDG~\cite{ParticleDataGroup:2024cfk}    &~~ $1713 \pm 19$ ~~ & ~~ $107 \pm 15$~~   \\
LHCb~\cite{LHCb:2023evz}    &~~ $1736 \pm 10 \pm 12 $ ~~ & ~~ $134 \pm 17 \pm 61$~~   \\
BESIII~\cite{BESIII:2022npc}   &~~ $1817 \pm 8 \pm 20$ ~~ & ~~ $97 \pm 22 \pm 15$~~   \\
Babar~\cite{BaBar:2021fkz}   &~~ $1704 \pm 5 \pm2$ ~~ & ~~ $110 \pm 15 \pm 11$~~   \\
Geng and Oset~\cite{Geng:2008gx}   &~~ $1777$ ~~ & ~~ $148$~~   \\
\hline\hline
\end{tabular}
\label{tab:1}
\end{table}

\begin{figure}[!bt]
	\begin{center}
	\includegraphics[scale=0.7]{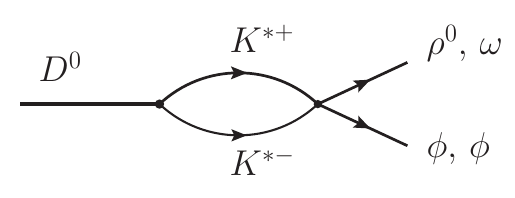}
	\end{center}
	\vspace{-0.7cm}
	\caption{Mechanism for $D^0 \to \rho^0 \phi, \, \omega \phi$ through $K^{*+}K^{*-}$ production and final state interaction.}
	\label{fig:Fig3}
\end{figure}

The mechanism for final state interaction after $K^{*+} K^{*-}$ production is given in Fig.~\ref{fig:Fig3}.
Analytically we have 
\begin{align}
	\label{eq:tL}
	t_L(\rho^0 \phi)=&A N_c \, G_{K^{*+}K^{*-}} (M_{D^0})\; T_{K^{*+}K^{*-},\, \rho^0 \phi},\nonumber\\
	t_L(\omega \phi)=&A N_c \, G_{K^{*+}K^{*-}} (M_{D^0})\; T_{K^{*+}K^{*-},\, \omega \phi},
\end{align}
where $A$ is a normalization constant that disappears in the ratios of decay widths and $N_c$ is the number of colors $N_c=3$. 
The same normalization constant $A$ will be used for internal emission, only the factor $N_c$ will not appear.
We take advantage to clarify that because the vertex $V_{q \bar q}$ is color blind, then in the external emission of Fig.~\ref{fig:2}, the color of the $s$ and $u$ quarks is the same but can be any of the three colors.
However, in the internal emission of Fig.~\ref{fig:1}, the $s$ quark has the color of the $c$ quark and the second $s$ quark has the color of the $u$ quark which are well defined in the $D^0$ or $\rho^0, \omega$ states. 
$G_{K^{*+}K^{*-}} (M_{D^0})$ in Eq.~\eqref{eq:tL} is the $K^{*+}K^{*-}$ loop function,
given in the cutoff regularization by
\begin{align}
	\label{eq:gi}
	G_{K^{*+}K^{*-}}(s)=\int_{\abs{\vec{q}}<q_\text{max}}\frac{\dd^3q}{(2\pi)^3}\frac{w_{1}(\vec{q}\,)+w_{2}(\vec{q}\,)}{2w_{1}(\vec{q}\,)w_{2}(\vec{q}\,)}\nonumber\\
	\times\frac{1}{s-[w_{1}(\vec{q}\,)+w_{2}(\vec{q}\,)]^2+i\epsilon},
\end{align}
with $w_{i}(\vec{q}\,)=\sqrt{\vec{q}\,^2+m_{i}^2}$, $m_i=m_{K^{*+}},m_{K^{*-}}$.
$T_{K^{*+}K^{*-},\, \rho^0 \phi}$ and $T_{K^{*+}K^{*-},\, \omega \phi}$ 
are the elements of the transition matrix $T$ from $K^{*+}K^{*-}$ to $\rho^0 \phi$ or $\omega \phi$ respectively.
These matrix elements are obtained in Ref.~\cite{Geng:2008gx} using the Bethe-Salpeter equation in coupled channels, 
\begin{equation}	\label{eq:BSeq}
	T=[I-VG]^{-1}V,
\end{equation}
with the potential $V_{ij}$ being provided by the local hidden gauge approach~\cite{Bando:1984ej,Bando:1987br,Meissner:1987ge,Nagahiro:2008cv}.
A value of $q_{\rm max}=960$~MeV was found suited in Refs.~\cite{Geng:2008gx,Dai:2021owu} to mach the dimensional regularization used there.  

In the $I=1$ sector of the $\rho \phi$ channel with spin $J=0$, the coupled channels are: $K^*\bar K^*, \rho \omega, \rho \phi$.
In the $I=0$ sector of $\omega \phi$ with $J=0$, the coupled channels are: $K^*\bar{K^*}, \rho \rho, \omega \omega, \omega\phi, \phi\phi$.
The amplitudes of the $T$ matrix of Eq.~\eqref{eq:BSeq} develop a pole, for $I=1$ the $a_0(1710)$, and for $I=0$ the $f_0(1710)$.
Close to the pole, the amplitudes behave as 
\begin{equation}
	\label{eq:Tij}
	T_{ij}=\dfrac{g_i\, g_j}{s-s_R},
\end{equation}
where $\sqrt{s_R}$ gives the position of the pole in the complex energy plane and $g_i$ is the coupling of the resonance to the coupled channel $i$.
The amplitudes of Eq.~\eqref{eq:Tij} can be taken in a range of energies about $2\Gamma$, with $\Gamma$ the width of the states, which means that for the decays that we are considering, the $T$ matrices can be well approximately by their pole form, which is actually a Breit-Wigner amplitude used in most experimental analyses of data.
Thus we have 
\begin{align}
 \label{eq:t_a0f0}
 T_{K^{*+} K^{*-},\, \rho^0 \phi}(s = M^2_{D^0}) &= \frac{g_{R,K^{*+} K^{*-}} \ g_{R,\rho^0 \phi} }{s - M_R^2 + iM_R \Gamma_R}, \nonumber\\
 T_{K^{*+} K^{*-},\, \omega \phi}(s = M^2_{D^0}) &= \frac{g_{R,K^{*+} K^{*-}} \ g_{R,\omega \phi} }{s - M_R^2 + iM_R \Gamma_R}.
\end{align}

The couplings in Eq.~\eqref{eq:t_a0f0} are given in Table~II ($I=1$) and Table~I ($I=0$) of Ref.~\cite{Geng:2008gx} and we reproduce them in Table~\ref{tab:2} here.
\begin{table*}[!tb]
\centering
\caption{Couplings of $f_0(1710)$ and $a_0(1710)$ to the needed coupled channels.}
\begin{tabular}{c| c c c}
\hline 
~~~~~ $a_0(1710)$ ($J=0$, $I=1$)~~~~~  &~~~$K^{*} \bar K^{*}$ ($I=1$) ~~ & ~~ $\rho \phi$ ~~  & ~~ $K^{*+} K^{*-}$~~~\\
 $g_i \; [{\rm MeV}]$   &~~ $7525 - i 1529$ ~~ & ~~ $4998 - i 1872$~~ &  $\displaystyle -\frac{1}{\sqrt{2}} g_{R,K^{*} \bar K^{*}(I=1)} $    \\
\hline 
 $f_0(1710)$ ($J=0$, $I=0$)  &~~~$K^{*} \bar K^{*}$ ($I=0$) ~~ & ~~ $\omega \phi$ ~~  & ~~ $K^{*+} K^{*-}$\\
 $g_i \; [{\rm MeV}]$  &~~ $7124 + i 96$ ~~ & ~~ $3010 - i 210$~~ &  $\displaystyle -\frac{1}{\sqrt{2}} g_{R,K^{*} \bar K^{*}(I=0)} $    \\
\hline
\end{tabular}
\label{tab:2}
\end{table*}

\begin{table*}[tb!]
	\centering
	\caption{Results for the ratio $R_b$ of Eq.~\eqref{eq:Rb} with different options.}
	 \label{tab:tabIII}
	\setlength{\tabcolsep}{18pt}
	\begin{tabular}{c|cccc}
	\hline
	\hline
	 &   Ref.~\cite{Geng:2008gx} &  Ref.~\cite{Geng:2008gx} $M_{a_0}=1750\, {\rm MeV}$ & PDG~\cite{ParticleDataGroup:2024cfk}& LHCb~\cite{LHCb:2023evz} \\[1.5mm]
	\hline
	$R_b$&$3.5 $&$ 2.8$&$ 2.2$&$ 2.5 $ \\
	\hline
	\hline
	\end{tabular}
\end{table*}

For the couplings of the resonance to $K^{*+} K^{*-}$, we have used the isospin wave function, with our isospin multiplets ($K^{*+}$, $K^{*0}$), ($\bar K^{*0}$, $-K^{*-}$).
\begin{align}
& | K^* \bar K^*, I=0 \rangle = \frac{1}{\sqrt{2}} \left( K^{*+} K^{*-}(-) - K^{*0} \bar K^{*0} \right), \\
& | K^* \bar K^*, I=1 \rangle = \frac{1}{\sqrt{2}} \left( K^{*+} K^{*-}(-) + K^{*0} \bar K^{*0} \right). 
\end{align}

The amplitudes for internal emission are given both by 
\begin{equation}
	\label{eq:tIE}
	t^{\rm IE}=\dfrac{1}{\sqrt{2}}\; A,
\end{equation}
where by putting the same normalization factor $A$ as in the loop mechanisms we are taking into account the $\displaystyle \frac{1}{N_c}$ reduction of internal emission versus external emission.
All this said, our amplitudes for $D^0$ decay are 
\begin{align}
	\label{eq:ti}
	t_{D^0, \, \rho \phi}=& \;t^{\rm IE}+t_L(\rho \phi),\nonumber\\
	t_{D^0, \, \omega \phi}=& \;t^{\rm IE}+t_L(\omega \phi),
\end{align}
and the decay widths will be given by 
\begin{equation}
	\label{eq:Width}
	\Gamma_i=\dfrac{1}{8\, \pi}\; \dfrac{1}{M^2_{D^0}}\; |t_i|^2\; q_i,
\end{equation}
with $q_i$ given by 
\begin{equation}
	q_{i}=\frac{\lambda^{1/2}(M^2_{D^0},m_\phi^2,{m_i'}^{2})}{2\, M_{D^0}},
\end{equation}
with $m'_i=m_\rho$ for $\rho \phi$ and $m'_i=m_\omega$ for $\omega \phi$ decay.
Formally $t_i$ in Eq.~\eqref{eq:Width} has dimension of energy, which means that the factor $A$ in Eq.~\eqref{eq:tL} has dimension of energy. Actually $A$ is unknown to us and this is why we evaluate the ratio of decay widths where this factor cancels out.
We define then the ratio of the widths as
\begin{equation}\label{eq:Rb}
	R_{b}=\frac{\Gamma_{D^0 \to \rho \phi}}{\Gamma_{D^0 \to \omega \phi}},
\end{equation}
and we obtain the results of Table \ref{tab:tabIII},
where we compare the results with the input of Ref.~\cite{Geng:2008gx},
with the PDG data~\cite{ParticleDataGroup:2024cfk} and with the input of the most recent experiment of LHCb~\cite{LHCb:2023evz}.
These results should be compared with the experimental ratio summing relative errors
\begin{equation}\label{eq:Rbexp}
	R^{\rm exp}_{b}=2.17\pm 0.52, ~ [1.65 - 2.69].
\end{equation}
We can see that the inclusion of the $K^{*+} K^{*-}$ production followed by rescattering to the final $\rho^0 \phi, \omega \phi$ states increases the rate of $D^0 \to \rho^0 \phi$ production versus that of $D^0 \to \omega \phi$ by about a factor $2$, as in the experiment.
We can see that using the result of Ref.~\cite{Geng:2008gx} one obtains a ratio $R_b$ $30\%$ bigger than the highest value of $R_b^{\rm exp}$ in the error band, which is not far away.
One reason for the extra enhancement is the fact that the mass of the $a_0$ is closer to $M_{D^0}$ than in Ref.~\cite{ParticleDataGroup:2024cfk}
or Ref.~\cite{LHCb:2023evz}, which makes the $K^{*+} K^{*-} \to \rho \phi$ amplitude of Eq.~\eqref{eq:t_a0f0} bigger. 

Since no uncertainties were given in Ref.~\cite{Geng:2008gx}, we take them from the paper of Ref.~\cite{Du:2018gyn}, with $m_{a_0} \in [1750-1790]$~MeV. 
If we take $m_{a_0} = 1750$~MeV, $R_b$ is reduced to $2.8$, more in agreement with the experimental value. 
The results obtained with the PDG or LHCb input are compatible with the experimental number. 
We could guess that the actual mass of the $a_0$ state could be closer to that of the LHCb measurement $1736 \pm 10 \pm 12$~\cite{LHCb:2023evz}, or the average given by the PDG of around $1713 \pm 19$~\cite{ParticleDataGroup:2024cfk}. 
Yet, with all the uncertainties, we prove the role of the $a_0(1710)$ state, increasing the $R_b$ ratio to reasonable numbers.

\section{Results with full amplitudes and $\rho$ mass distribution}
\subsection{Full amplitudes in coupled channels}
In Eq.~\eqref{eq:t_a0f0} we approximated the amplitudes needed in our approach by Breit-Wigner forms taking the couplings used in Ref.~\cite{Geng:2008gx}. Here we repeat the calculations of Ref.~\cite{Geng:2008gx} and approximate the contribution of the box diagrams, which have relevance on the decay of the resonances to two pseudoscalars, but small repercussion on the mass of these resonances. 

The potentials are given in Appendix A of the arXiv version of Ref.~\cite{Geng:2008gx}, and one has
\begin{equation}\label{eq:Vex}
	V=V_{c}+V_{ex},
\end{equation}
where $V_{c}$ is the contact term, and $V_{ex}$ the one coming from vector exchange. We have:
\begin{itemize}
	\item[1)] $S=0, I=0, J=0 \;(f_0(1710))$. The coupled channels are
	$K^{*} \bar K^{*}$, $\rho \rho$, $\omega \omega$, $\omega \phi$, $\phi \phi$. The contact terms, $V_{\rm c}$, are taken from Table V of Ref.~\cite{Geng:2008gx} and the vector exchange terms from Table XVIII.

	The box diagrams are very difficult to calculate (see Eq.~(C1) of Ref.~\cite{Geng:2008gx}), but we find an easy way to circumvent this problem which has been successfully tried in similar problems, like to take into account decay channels of a $D\bar D$ bound state (see Ref.~\cite{Dai:2015bcc,Wang:2019evy,Dai:2020yfu,Wang:2020elp}).
	The strategy is the following. 
	We introduce a pseudoscalar-pseudoscalar ($PP$) decay channel and couple it to the dominant $VV$ channel ($K^* \bar K^*$ in the present case), and tune the transition amplitude $K^* \bar K^* \to PP$ to get the width obtained in Ref.~\cite{Geng:2008gx}.
	For this we choose also the dominant $PP$ channel in the decay of the resonance, the $K \bar K$.
	Then introduce the $K\bar K$ as a new coupled channel assuming
\begin{equation}\label{eq:NewVij1}
\begin{aligned}
    &V_{K^* \bar K^*, K\bar K} = a, \\[2mm]
    &V_{K\bar K, K\bar K} = 0, \\[2mm]
    &V_{VV, K\bar K} = 0,~~{\rm for~ any~} VV \neq K^* \bar K^*.
\end{aligned}
\end{equation}
Also, since the $K\bar K$ channel is far away from $K^* \bar K^*$, we take the $K\bar K$ loop function as
\begin{equation}\label{eq:GKK0}
	G_{K\bar K} (M_{\rm inv}) =i\; \Im G_{K\bar K}(M_{\rm inv})=-i \; \dfrac{1}{8\pi \, M_{\rm inv}}\; q_k,
\end{equation}
with 
\begin{equation}\label{eq:GKK}
	q_k=\dfrac{\lambda^{1/2}(M^2_{\rm inv}, m^2_K, m^2_K)}{2 \,M_{\rm inv}}.
\end{equation}

In Eq.~\eqref{eq:GKK0} we take $\Re G_{K\bar K}=0$ since the channel is very far away from the $K^* \bar K^*$ threshold and $G_{K\bar K}$ is largely dominated by its imaginary part.
With this set up, we tune $q_{\rm max}$ and the parameter $a$ to get the mass and width of the $f_0(1710)$ state obtained in Ref.~\cite{Geng:2008gx}.
This is accomplished by solving the Bethe-Salpeter equation of Eq.~\eqref{eq:BSeq} with the enlarged base of coupled channels, and choosing
\begin{equation*}
	q_{\rm max}=980\; {\rm MeV}, ~~~~~a=55,
\end{equation*}
and the pole oppears at $(1725.25+i 65.28)\; {\rm MeV}$.
Then we take
\begin{equation}\label{eq:Newb}
	T_{K^{*+}K^{*-}, \,\omega \phi}=-\dfrac{1}{\sqrt{2}}\, T_{K^{*}\bar K^{*},\, \omega \phi}.
\end{equation}

\item[2)] $S=0, I=1, J=0 \;(a_0(1710))$. The coupled channels are
$K^{*} \bar K^{*}$, $\rho \omega$, $\rho \phi$. We take the potential of Eq.~\eqref{eq:Vex}, with $V_{\rm c}$ and $V_{\rm ex}$ taken from Table VIII and Table XX of Ref.~\cite{Geng:2008gx} respectively. 

We proceed as before introducing the $K\bar K$ channel and coupling it to the main $K^* \bar K^*$ component as in Eq.~\eqref{eq:NewVij1} but taking now
\begin{equation}
	V_{K^* \bar K^*, K\bar K}=b.
\end{equation}
Once again we solve the Bethe-Salpeter equation with the enlarged base of coupled channels and tune $q_{\rm max}$ and $b$ to get the mass and width of the $a_0$ state as found in Ref.~\cite{Geng:2008gx}.
This is accomplished by choosing
\begin{equation*}
	q_{\rm max}=980\; {\rm MeV}, ~~~~~b=15,
\end{equation*}
which leads to the pole at $(1779.33+i 85.05)\; {\rm MeV}$.
Then we take
 \begin{equation}
	T_{K^{*+}K^{*-},\, \rho^0 \phi}=-\dfrac{1}{\sqrt{2}}\, T_{K^{*}\bar K^{*}, \,\rho \phi}.
\end{equation}
\end{itemize}
With all these ingredients, we obtain now a ratio
\begin{equation}\label{eq:Rbnew}
 R_b^{\rm new}=1.97.
\end{equation}

\subsection{Convolution of $\Gamma_{D^0 \to \rho \phi}$ with the $\rho$ mass distribution}

In the $D^0 \to \rho \phi$ decay, there is only about $78 \, \rm MeV$ of excess energy, but the $\rho$ has a width of $\Gamma_\rho =147.4 \, \rm MeV$. Then, the consideration of the $\rho$ mass distribution should be relevant in the $D^0 \to \rho \phi$ decay. To take this into account, we change Eq.~\eqref{eq:Width} to
\begin{eqnarray} \label{eq:Gamma}
  \Gamma_{D^0 \to \rho \phi}&=& - \frac{1}{\pi}\int d \tilde{m}^2_\rho \;\Im \dfrac{1}{\tilde{m}^2_\rho-m_\rho^2 +i \tilde{m}_\rho \, \Gamma_\rho}\nonumber \\
&& \times \, \frac{1}{8\pi} \, \frac{1}{m_{D^0}^2}\, \left|t_{D^0, \, \rho \phi}\right|^2 \, \tilde{q}_\rho,
\end{eqnarray}
with 
\begin{equation}
	\tilde{q}_\rho=\dfrac{\lambda^{1/2}(M^2_{D^0},m_\phi^2,\tilde{m}^2_\rho)}{2\, M_{D^0}}.
\end{equation}
With all these changes, we obtain now a ratio
\begin{equation}\label{eq:Rbnew}
 R_b^{\rm Con}=1.47.
\end{equation}
As we can see, the results are in line with those obtained in Table \ref{tab:tabIII}, although somewhat smaller, and in line with the experimental values of Eq.~\eqref{eq:Rbexp}.

\section{Conclusions}
We have studied the $D^0 \to \rho^0 \phi$, $\omega \phi$ decays, which produce directly $\rho^0 \phi$, $\omega \phi$ with the internal emission mechanism with equal rates. 
However, experimentally the rate for $\rho^0 \phi$ production is about double the one of $\omega \phi$ production. We found a natural explanation of this ratio resorting to an indirect production mode, where $K^{*+} K^{*-}$ is produced in external emission in a first step, and the $K^{*+} K^{*-}$ undergoes final state interaction with coupled $VV$ channels to give $\rho^0 \phi$ in $I=1$, or $\omega \phi$ in $I=0$.
The $VV$ interaction in these $VV$ channels produces two resonances, the $f_0(1710)$ in $I=0$, well known in the literature, and the $a_0(1710)$ that was predicted in Ref.~\cite{Geng:2008gx} in 2009 and was only found experimentally in 2021 by the Babar collaboration, followed by two other experiments in BESIII in 2022 and LHCb in 2023.
We found that mostly due to the stronger coupling of $a_0(1710)$ to the $\rho \phi$ channel than the $f_0(1710)$ to the $\omega \phi$ channel, the indirect production of $K^{*+}K^{*-}$ followed by $f_0(1710)$ or $a_0(1710)$ production to lead to $\omega \phi$ or $\rho^0 \phi$ respectively, was responsible for an appreciable increase of the $\rho \phi$ to the $\omega \phi$ rate, in fair agreement with experiment.
While the role of this newly found $a_0(1710)$ state has manifested itself clearly in the reactions, future measurements with better precision can help us determine better its properties, mostly concerning its mass, where there are still some differences among the different experiments.

\section*{Acknowledgments}
We warmly thank Prof. Qiang Zhao for calling us the attention to this problem and useful discussions.
And we thank Prof. Chu-Wen Xiao and Dr. Hai-Peng Li for useful discussions.
N. I. would like to express gratitude to Guangxi Normal University for their warm hospitality, as part of this work was conducted there. 
This work is partly supported by the National Natural Science Foundation of China (NSFC) under Grants No. 12365019, and No. 12575081, 
and by the Central Government Guidance Funds for Local Scientific and Technological Development, China (No. Guike ZY22096024), 
and by the Natural Science Foundation of Guangxi province under Grant No. 2023JJA110076.
This project has received funding from the European Union Horizon 2020 research and innovation programme under the program H2020-INFRAIA-2018-1, grant agreement No. 824093 of the STRONG-2020 project.
This work of N. I. was partly supported by JSPS KAKENHI Grant Number 24K07020.

\bibliography{ref_D0decay}

\end{document}